\documentclass[9pt,twocolumn,twoside]{pnas-new}

\graphicspath{{./Figs/}}

\templatetype{pnasresearcharticle} 

\title{Variational implicit-solvent predictions of the dry-wet transition pathways 
for ligand-receptor binding and unbinding kinetics}

\newcommand{\br}{\mathbf{r}}            

\newcommand{\ve}{\varepsilon}           

\newcommand\redout{\bgroup\markoverwith
{\textcolor{red}{\rule[.5ex]{2pt}{0.4pt}}}\ULon}

\author[a]{Shenggao Zhou}
\author[b]{R. Gregor Wei\ss} 
\author[c]{Li-Tien Cheng}
\author[d]{Joachim Dzubiella}
\author[e,1]{J. Andrew McCammon}
\author[c,1]{Bo Li}

\affil[a]{Department of Mathematics and Mathematical Center for Interdiscipline Research, 
          Soochow University, 1 Shizi Street, Suzhou 215006, Jiangsu, China}
\affil[b]{Laboratory of Physical Chemistry, ETH Z{\"u}rich, 
          Vladimir-Prelog-Weg 2, CH-8093 Z{\"u}rich, Switzerland; 
          and Institut f{\"u}r Physik, Humboldt-Universit{\"a}t zu Berlin, 
          Newtonstrasse 15, D-12489 Berlin, Germany}
\affil[c]{Department of Mathematics, University of California, San Diego, 
          9500 Gilman Drive, La Jolla, California 92093-0112, USA}
\affil[d]{Physikalisches Institut, Albert-Ludwigs-Universit{\" a}t 
          Freiburg, Hermann-Herder-Stra{\ss}e 3, 79104 Freiburg, Germany; 
          and Research Group Simulations of Energy Materials (EE-GSEM), 
          Helmholtz-Zentrum Berlin, Hahn-Meitner-Platz 1, 14109, Berlin, Germany}
\affil[e]{Department of Chemistry and Biochemistry, Department of Pharmacology, 
          University of California, San Diego, 
          9500 Gilman Drive, La Jolla, California 92093-0365, USA}

\leadauthor{Zhou}

\significancestatement{

The kinetics of ligand-receptor (un)binding---how fast a ligand binds into and resides in 
a receptor---cannot be inferred solely from the binding affinity which describes the 
thermodynamic stability of the bound complex. A bottleneck in understanding such kinetics, 
which is critical to drug efficacy, lies in the modeling of the collective water fluctuations in 
apolar confinement. We develop a new theoretical approach that couples a variational 
implicit-solvent model with the string method to describe the dry-wet transition pathways, 
which then serve as input for the ligand 
multi-state Brownian dynamics.
Without explicit descriptions of individual water molecule,
our theory predicts the key 
thermodynamic and kinetic 
properties of unbinding and binding, the latter in quantitative agreement with 
explicit-water molecular dynamics simulations.
}

\authorcontributions
{
JD, JAM, and BL designed research.  
SZ, RGW, and LTC performed research. 
SZ, LTC, and BL developed numerical methods and LTC wrote the initial level-set code. 
SZ, RGW, and JD analyzed computational results.     
SZ, RGW, JD, JAM, and BL wrote the paper. 
}
\authordeclaration{The authors declare no conflict of interest.}
\correspondingauthor{\textsuperscript{1}To whom correspondence should be addressed. 
E-mail: jmccammon@ucsd.edu or bli@math.ucsd.edu}

\keywords{Ligand-receptor binding/unbinding kinetics $|$ dry-wet transitions $|$ 
variational implicit-solvent model  $|$ level-set method $|$ string method} 

\begin{abstract}
Ligand-receptor binding and unbinding are fundamental biomolecular processes and particularly
essential to drug efficacy. Environmental water fluctuations, however,  
impact the corresponding thermodynamics and kinetics  and thereby challenge theoretical descriptions.  
Here, we  devise a holistic, implicit-solvent, multi-method approach 
to predict the (un)binding kinetics for a generic ligand-pocket model. 
We use the variational implicit-solvent model (VISM) to calculate the 
solute-solvent interfacial structures and the corresponding free energies, 
and combine the VISM with the string method to obtain  
the minimum energy paths and transition states between the 
various metastable (``dry'' and ``wet'') hydration states.  
The resulting dry-wet transition rates are then used in a 
spatially-dependent multi-state continuous-time Markov chain Brownian dynamics 
simulations, and the related Fokker--Planck equation calculations, 
of the ligand stochastic motion, providing the mean first-passage times for binding and unbinding.
We find the hydration transitions to significantly slow down the binding process, 
in semi-quantitative agreement with
existing explicit-water simulations, but significantly accelerate the unbinding process.
Moreover, our methods allow the characterization of non-equilibrium 
hydration states of pocket and ligand during the ligand movement, 
for which we find substantial memory and hysteresis effects for binding versus unbinding.
Our study thus provides a significant step forward towards efficient, 
physics-based interpretation and predictions of the complex kinetics in realistic ligand-receptor systems. 

\end{abstract}

\dates{This manuscript was compiled on \today}
\doi{\url{www.pnas.org/cgi/doi/10.1073/pnas.XXXXXXXXXX}}

\begin{document}

\maketitle
\thispagestyle{firststyle}
\ifthenelse{\boolean{shortarticle}}{\ifthenelse{\boolean{singlecolumn}}{\abscontentformatted}{\abscontent}}{}



\dropcap{T}he complex process of ligand-receptor binding and unbinding in aqueous 
environment is fundamental to biological function. Understanding the 
thermodynamics and kinetics of such processes has far-reaching practical 
significance, particularly in rational drug design~\cite{Babine_ChemRev1997, Shaw_Today2013}. 
Water is a key player in ligand-receptor binding and unbinding, and 
in molecular recognition in general 
\cite{
LevyOnuchic_Rev06,
BaronMcCammon_Rev13}.  
In particular, it has been well established that hydrophobic interactions
can drive the association and dissociation of biological molecules
\cite{
Pratt_Rev_ChemPhys2000,
Chandler05, 
BerneWeeksZhou_Rev09, 
Ben-Amotz2016}.


Hydration contributes significantly to the ligand-receptor binding free energy, 
determining the thermodynamic stability of the bound unit
\cite{
McCammon_BiophysJ1997,
GilsonZhou_Rev2007}. 
Recent experimental and theoretical studies have indicated that the kinetics of ligand-receptor
binding and unbinding is 
crucial for drug effectiveness and efficacy
\cite{Shaw_Today2013,
Bernetti_MedChemComm2017,
Ecker_DrugDiscToday2017}.  
Often, a ligand binds to a hydrophobic pocket on the surface of a receptor molecule 
\cite{
 YoungBF_PNAS2007,
 Halle:PNAS2008,
 WBFriesner_PNAS11, 
Sherman_Proteins2012}. 
Water molecules fluctuate around such an apolar pocket,
leading to metastable ``dry'' or ``wet'' hydration states of the binding site,
separated by an energetic barrier which is on the order of $k_{\rm B} T$~\cite{VISM_LS_PRL09}.
Such a moderate energetic hurdle facilitates repeated condensation and
evaporation of water in the pocket region, leading 
to large collective hydration fluctuations~\cite{SMJD_PNAS13}.  
In general, the dewetting of local regions generates strong 
hydrophobic forces in molecular association and dissociation
\cite{
HMBerne_PNAS03,
Chandler05, 
Berne_Nature2005,
BerneWeeksZhou_Rev09}. 
In particular, it has been demonstrated that the dry-wet transitions are
a precursor of the ligand-receptor binding and unbinding
\cite{
VISM_LS_PRL09,
BaronSetnyMcCammon_JACS10, 
Weiss:JCTC2017}.
Besides being the origin for the thermodynamically driven forces, 
water fluctuations also modify the friction and kinetics of associating hydrophobic molecules
\cite{
MorroneLiBerne_JPCB2012,
LiMorroneBerne_2012,
JJBerne_PNAS13,
UbdBerne_PNAS15,
Weiss:JPCB2016}, 
slowing down the binding kinetics and giving rise to local non-Markovian 
effects~\cite{SMJD_PNAS13, Weiss:JPCB2016}.

While water plays a critical role in molecular recognition,
efficient modeling of water is rather challenging
due to an overwhelming number of solvent degrees of freedom, many-body effects, and the
multi-scale nature of molecular interactions.
Explicit-water molecular dynamics (MD) simulations have been the main tool in
most of the existing studies of the kinetics of 
ligand-receptor binding and unbinding
\cite{
Shaw_BindingPathway_PNAS2011,
SMJD_PNAS13,
JJBerne_PNAS13,
Noe_NatCommun2015,
Parrinello_PNAS2015,
UbdBerne_PNAS15,
Weiss:JCTC2017,
Amaro_JPCL2018,
Miao_Biochem2018,  
Wade_BindingKinetics_2018}. 
While explicitly tracking water molecules,
MD simulations are still limited to
systems of relatively small sizes and events of relatively short time scales. 
In particular, slow and rare water fluctuations and large ligand residence 
times in the pocket still challenge the prediction of unbinding times.


In this work, we develop a holistic, multi-method, implicit-solvent approach to study
the kinetics of ligand-receptor binding and unbinding 
in a generic pocket-ligand model exactly as studied previously 
by explicit-water MD simulations~\cite{SMJD_PNAS13},
focusing on the effect of solvent fluctuations and multiple hydration states on such processes.  


Our approach is based on the variational implicit-solvent model (VISM) 
that we have developed in recent years \cite{DSM06a,DSM06b,CDML_JCP07,
WangEtal_VISMCFA_JCTC12,VISMPB_JCTC2014}. 
In VISM, one minimizes a 
solvation free-energy functional of solute-solvent interfaces
to determine a stable, equilibrium conformation, and to provide an approximation of the solvation
free energy.  The functional couples the solute surface energy, solute-solvent
van der Waals (vdW) dispersive interactions, and electrostatics.
This theory resembles that of Lum--Chandler--Weeks \cite{LCW99}   
[cf.\ also 
\cite{
Borgis_JPCB2005,BatesEtal_JMB09}],  
and is different from the existing SAS (solvent-accessible surface) type 
models.  We have designed and implemented a robust level-set method to 
numerically minimize the VISM functional with arbitrary 3D geometry 
\cite{CDML_JCP07,
ChengLiWang_JCP10,
WangEtal_VISMCFA_JCTC12,
VISMPB_JCTC2014}.


Here, for our model ligand-pocket system, 
we use our level-set VISM to obtain different hydration states and their 
solvation free energies, and use the VISM-string method \cite{ERE_JCP07, RE_JCP13}
to find the minimum energy paths connecting such states
and the corresponding transition rates.  Such rates are then used in our 
continuous-time Markov chain Brownian dynamics simulations, 
and the related Fokker--Planck equation calculations, of the ligand stochastic motion 
to obtain the mean first-passage times for the ligand binding and unbinding.  
We compare our results with existing explicit-water MD simulations.

\subsection*{The model ligand-receptor system}

\begin{figure}[th]
\centerline{\includegraphics[width=3.5in,height=1.3in]{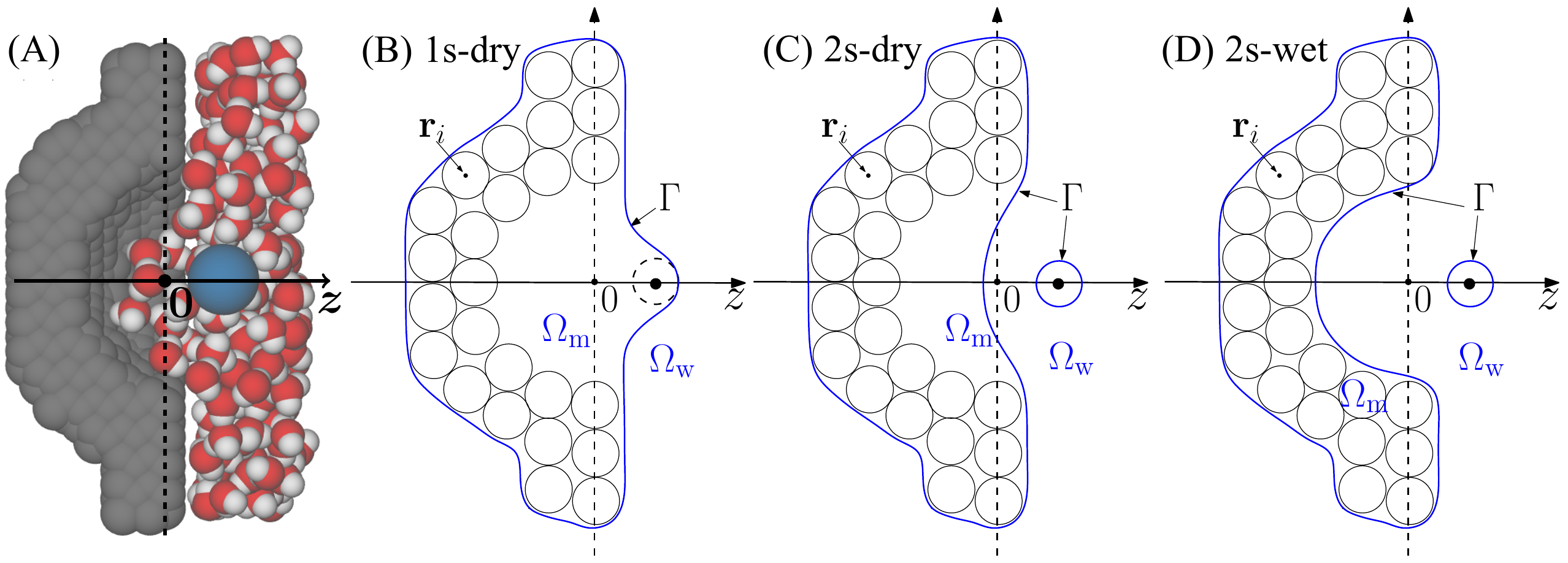}}
\caption{
(A) 
A schematic of the ligand (blue sphere),
explicit water, and the pocket of a concave wall.
(B) 1s-dry: The VISM surface (blue line) 
is a single surface enclosing all the wall atoms and also the ligand atom, 
hence a dry state of the pocket. 
(C) 2s-dry: The VISM surface has two disjoint components, one enclosing
all the wall atoms with a dry pocket, and one enclosing the ligand. 
(D) 2s-wet: The VISM surface has two components, tightly wrapping up the wall and
ligand, respectively, with no space for water, 
hence a wet pocket.
}
\label{f:geometry}
\end{figure}

The generic pocket-ligand model~\cite{SetnyGeller_JChemPhys06} consists of a hemispherical 
pocket and a methane-like molecule; cf.\ Fig.~\ref{f:geometry} (A).
The pocket, 
with the radius $R = 8$ {\AA} and centered at $(0,0,0)$,
is embedded in a rectangular wall, composed of apolar atoms
aligned in a hexagonal close-packed grid of lattice constant $1.25$ {\AA}.
The wall surface is oriented in $xy$-plane.
The ligand, a single neutral 
Lennard-Jones (LJ) sphere, 
is placed along the pocket symmetry axis, the $z$-axis, 
which is taken to be the reaction coordinate. 
Fig.~\ref{f:geometry} (B)--(D) depict the cross sections of all the possible 
VISM surfaces, i.e., the stable solute-solvent interfaces separating the solute region $\Omega_{\rm m}$ 
and solvent region $\Omega_{\rm w}$, representing different hydration states for a fixed 
position of ligand.



\section*{Results and Analysis}
\label{s:Results}


\subsection*{Multiple hydration states and the potential of mean force (PMF)}

We use our level-set method to minimize the VISM 
solvation free-energy functional (cf.\ Eq.~[\ref{G}] in Theory and Methods) and obtain
a VISM surface.
By choosing different initial solute-solvent interfaces, we obtain different VISM surfaces
describing different hydration states; cf.\ Fig.~\ref{f:geometry}.    

\begin{figure}[h]
\centerline{\includegraphics[width=2.5in,height=1.9in]{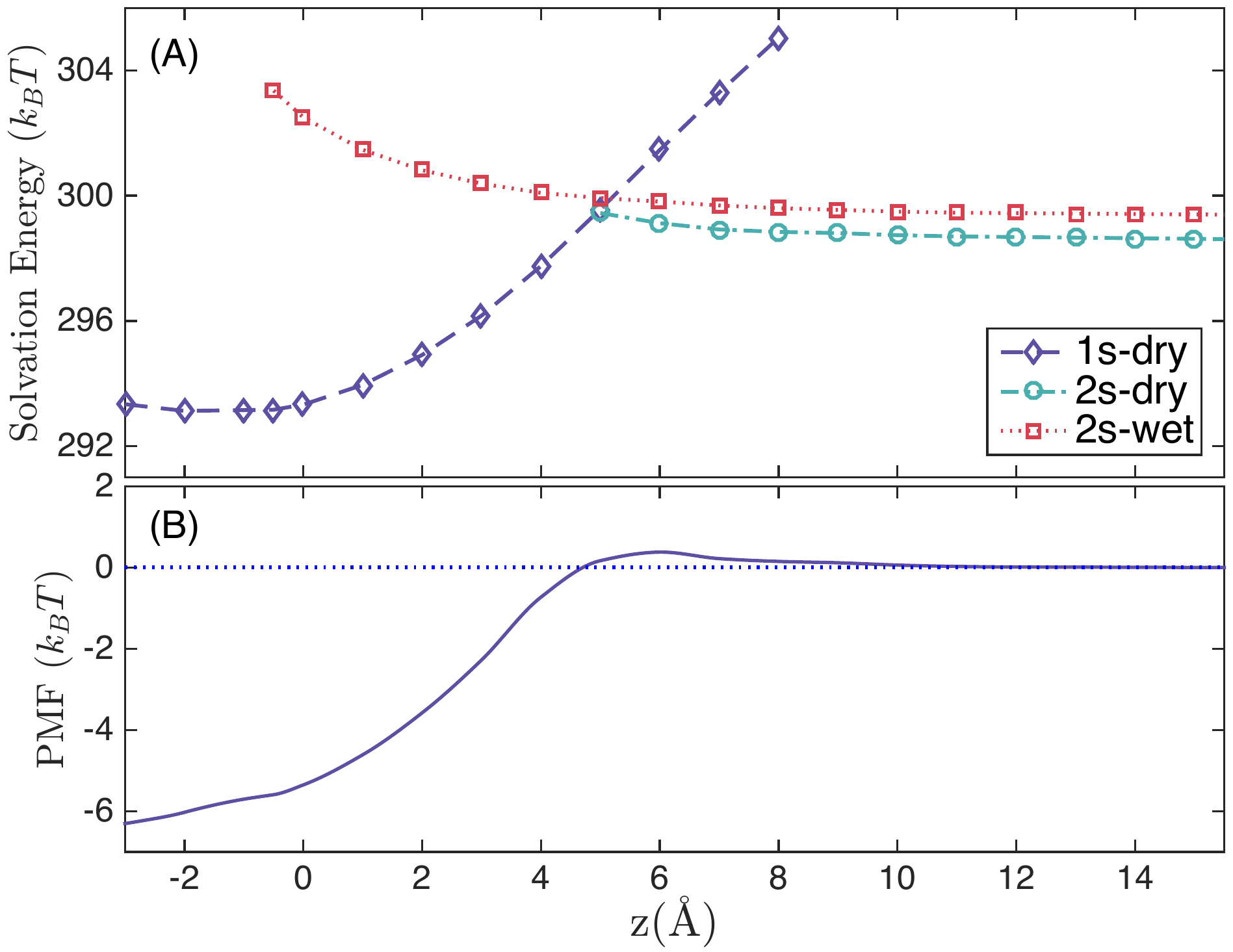}}
\caption{
(A) Solvation free energies of different VISM surfaces vs.\ the ligand location. 
(B) The equilibrium PMF.
}
\label{f:SolEngy}
\end{figure}

Fig.~\ref{f:SolEngy} (A) shows the solvation free energies for different VISM surfaces
against the reaction coordinate $z$.  For  $z < -0.5$ {\AA}, 
there is only one VISM surface, 1s-dry;
 cf.\ Fig.~\ref{f:geometry} (B). 
In addition to 1s-dry, a second VISM surface, 2s-wet, appears for 
$-0.5 < z <  5$ {\AA}; 
 cf.\ Fig.~\ref{f:geometry} (D). 
For $5 < z < 8 $ {\AA}, there are three VISM surfaces. In addition
to 1s-dry and 2s-wet, the third one is 2s-dry; 
cf.\ Fig.~\ref{f:geometry} (C).
Once the ligand is away from the pocket with $ z > 8$ {\AA}, there 
are only two VISM surfaces: 2s-dry and 2s-wet.

Fig.~\ref{f:SolEngy} (B) shows the equilibrium PMF, defined as
\begin{equation}
\label{V}
V(z)=- k_{\rm B}T \ln \biggl( { \sum_{\Gamma (z)} 
e^{-G[\Gamma (z) ]/k_{\rm B}T}} \biggr) +U_0(z) + V_{\infty},
\end{equation}
where $\Gamma(z)$ runs over all the VISM surfaces with $G[\Gamma(z)]$ the VISM
solvation free energy at $\Gamma(z)$, and  
$ U_0(z) = \sum_{\br_i} U_{\rm LJ} (| \br_i - \br_z |)$ with $\br_z$ the ligand position vector, 
$\br_i$ running through all the wall atoms, and $U_{\rm LJ}(r)$ a $12$--$6$ LJ potential. 
The constant $V_{\infty} $ is chosen so that $V(\infty) = 0.$
The PMF agrees well with the result from 
MD simulations \cite{Setny_JChemPhys07, CWSDLM_JCP09, VISM_LS_PRL09}.  
\subsection*{Dry-wet transition paths and energy barriers} 

At a fixed reaction coordinate $z$ with multiple hydration states, 
we use our level-set VISM coupled with the string method to calculate the minimum energy
paths (MEPs) that connect these states, and the corresponding transition states, energy barriers, 
and ultimately the transition rates.  A string or path here
consists of a family of solute-solvent interfaces,
and each point of a string, which is an interface in our case,  
is called an image.

\begin{figure}[h]
\centerline{\includegraphics[width=3.1in,height=2.4in]{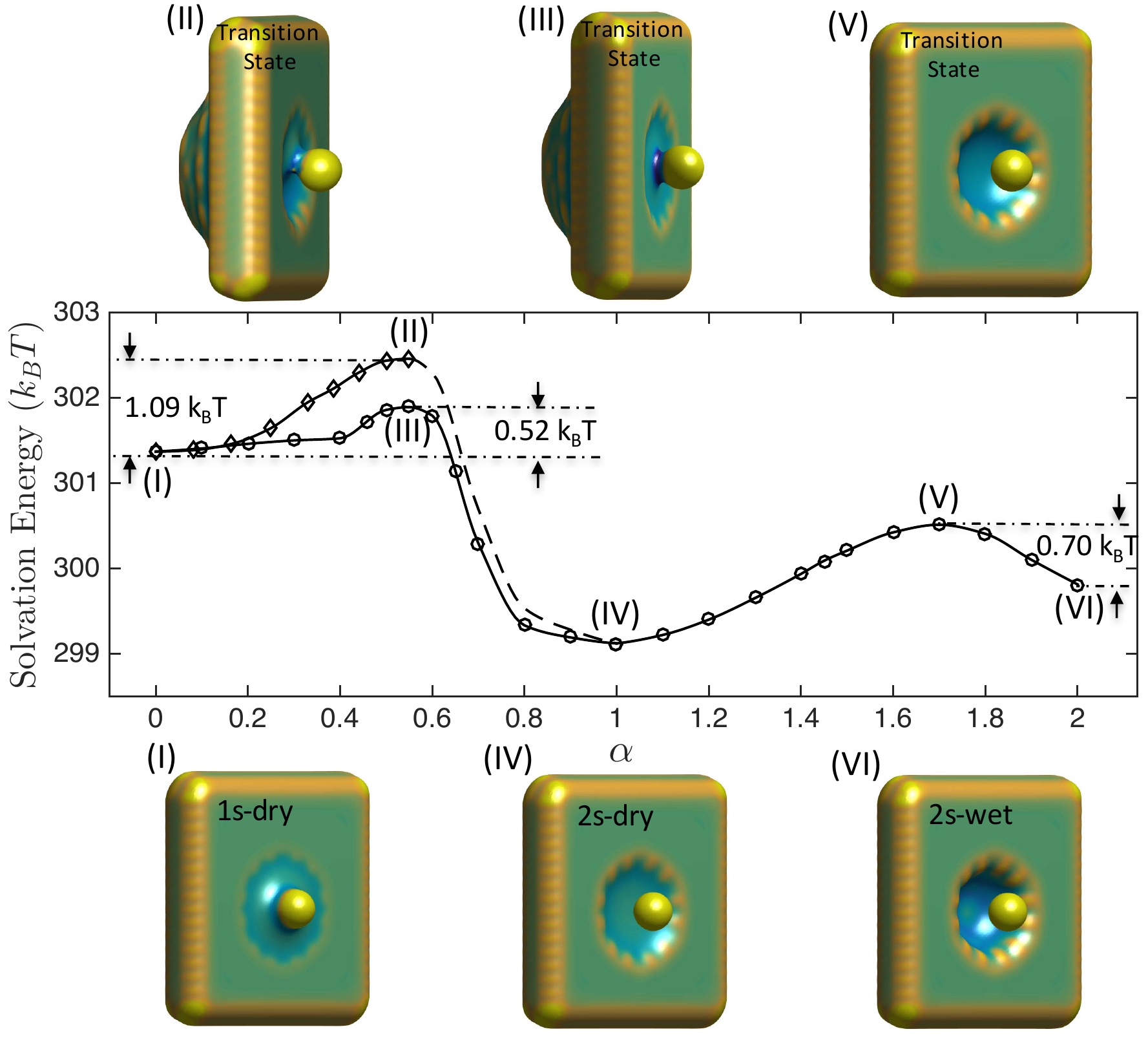}}
\caption{Solvation free energies of images on MEPs
that connect the hydration states 1s-dry (I), 2s-dry (IV), and 2s-wet (VI) 
(shown in the bottom) with transition states (II), (III), and (V) (shown on top) and 
the transition energy barriers
for $z = 6$ {\AA}.
In the middle plots, the horizontal axis is the string parameter $\alpha.$  
}
\label{f:d6}
\end{figure}

In Fig.~\ref{f:d6}, we display the solvation free energies of images
on MEPs that connect the three hydration states, 1s-dry, 2s-dry, and 2s-wet, at $z = 6$ {\AA}. 
There are two MEPs connecting 1s-dry (marked (I)) and 2s-dry (marked (IV)). 
One of them passes  
through the axisymmetric transition state marked (III),
and the other 
passes through the axiasymmetric transition state marked (II). 
Here, symmetry or asymmetry refers to that of the 3D conformation of the VISM surface.
Energy barriers in the transition from the state 1s-dry to 2s-dry along the two transition 
paths are estimated to be $1.09$ $k_{\rm B} T $ and $0.52$ $k_{\rm B}T$, respectively. 
Only one MEP is found to connect  2s-dry (marked (IV)) and 2s-wet (marked (VI)), 
and the corresponding transition state (marked (V)) is also found.
The MEP from 1s-dry to 2s-wet always passes through the state 2s-dry. 

\begin{figure}[h]
\centerline{\includegraphics[scale=0.38]{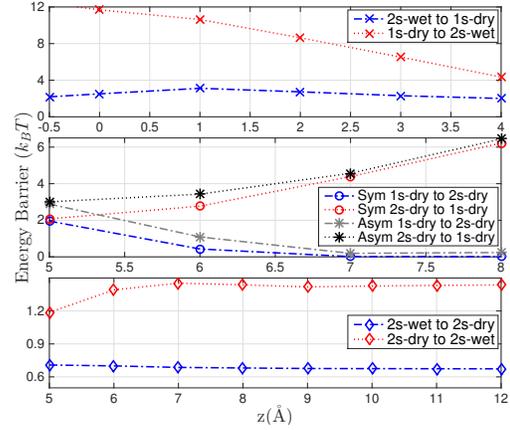}}
\caption{Transition energy barriers vs.\ the reaction coordinate $z$ 
with $-0.5 \le z \le 4$ {\AA} (top) and  
$5 \le z \le 12$ {\AA} (middle and bottom).   
Sym or Asym stands for a MEP with an axisymmetric or axiasymmetric transition state. 
}
\label{f:BarrEngy}
\end{figure}

Fig.~\ref{f:BarrEngy} summarizes all the energy barriers in the transitions 
from one hydration state to another for each reaction coordinate $z.$
For $0 \le z \le 4 $ {\AA} shown in the top of Fig.~\ref{f:BarrEngy}, 
there are only two hydration states: 1s-dry and 2s-wet. The 1s-dry has 
a lower free energy; cf.\ Fig.~\ref{f:SolEngy} (A), and hence
the barrier in the wetting transition from 1s-dry to 2s-wet (shown in red) is higher  
than that in the dewetting transition from 2s-wet to 1s-dry (shown in blue).
The dewetting barrier first increases as the ligand approaches 
the entrance of the pocket (from $z=4 $ to  $ z = 1 $ {\AA}), and then decreases after the ligand 
enters the pocket (from $z=1$ to $z = -0.5 $ {\AA}). 
This is because that more attractive solute-solvent 
vdW interaction is lost in dewetting as the ligand-pocket distance reduces from $z = 4$ to 
$z = 1$ {\AA}, and that the decrease in interfacial energy outweighs the vdW contribution 
to the solvation free energy as the distance further reduces from $z = 1$ to $z = -0.5$ {\AA}.  
Our predictions agree well with those by the explicit-water MD simulations~\cite{VISM_LS_PRL09}.  

For $5 \le z \le 8$ {\AA}, there are three hydration states 1s-dry, 2s-wet, and 2s-dry;  
cf.\ Fig.~\ref{f:SolEngy} (A). 
In the middle of Fig.~\ref{f:BarrEngy}, we plot for $z$ in this range the energy barriers 
along the MEPs, both axisymmetric 
and axiasymmetric, 
connecting the two states 1s-dry and 2s-dry; cf.\ Fig.~\ref{f:d6}. 
Note that, as the ligand approaches the pocket, the solute-solvent interfacial energy changes rapidly, 
and hence the barrier in the transition from 1s-dry to 2s-dry 
increases quickly, while the barrier in the reverse transition decreases quickly.  

In the bottom of Fig.~\ref{f:BarrEngy}, we plot energy barriers
for transitions between the states 2s-dry and 2s-wet
in the range $5 \le z \le 12$ {\AA}; cf.\ Fig.~\ref{f:SolEngy} (A). 
As the ligand-pocket distance increases, the barrier for the wetting transition (marked red)
first increases, since the newly created solvent region with attractive solute-solvent vdW
interaction decreases.  It then reaches a plateau after the distance is greater than $7$ {\AA.}  
The pocket dewetting barrier (marked blue) is slightly larger 
when the ligand is close to the pocket, since 
contributions of solute-solvent vdW interaction are lost during the pocket dewetting.

\subsection*{Kinetics of binding and unbinding}


We perform continuous-time Markov chain (CTMC) 
Brownian dynamics (BD) simulations and solve the related Fokker--Planck equation (FPE) calculations 
for the ligand stochastic motion with the pocket dry-wet fluctuations; 
see Theory and Methods.
For comparison, we also perform the usual BD
simulations and FPE calculations without including such fluctuations. 

\begin{figure}[ht]
\centerline{
\includegraphics[width=0.47\textwidth]{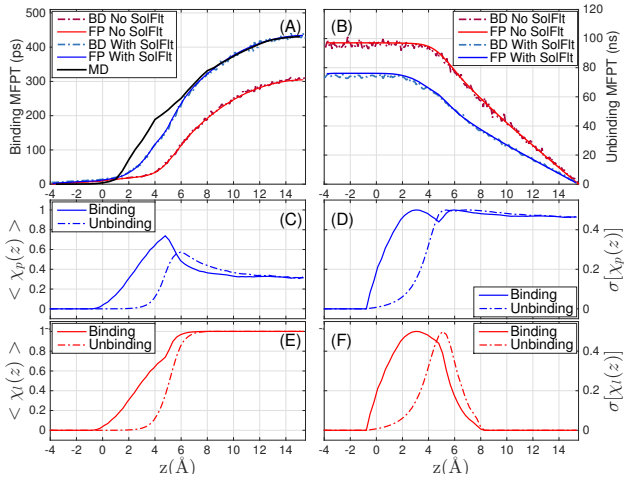}
}
\caption{
The MFPT for: (A) the binding of ligand that starts
from $z_{\rm init} = z$ and reaches the pocket at $z_{\rm L} = - 4$ {\AA};  
and (B) the unbinding of ligand that starts from  $z_{\rm init} = z$ and reaches 
$z_{\rm R} = 15.5$ {\AA}, predicted by: BD simulations without (BD No SolFlt) and with 
(BD With SolFlt) the dry-wet fluctuations; and FPE calculations without (FP No SolFlt) 
and with (FP With SolFlt) the dry-wet fluctuations, respectively. 
Note that the time unit on the vertical axis in (B) is ns while that in (A) is ps. 
The MFPT obtained by explicit-water MD simulations (MD)~\cite{SMJD_PNAS13} is also shown in (A). 
(C)--(F) The mean values and standard deviations of the pocket and ligand hydration 
states $\chi_p(z)$ and $\chi_l(z)$, respectively, against the ligand location $z$ 
during the {\it nonequilibrium} binding process 
from the BD simulations starting at $z_{\rm init}=6$~{\AA} (cf.\ (C) and (E)) and 
the unbinding process starting at $z_{\rm init}=-2$~{\AA} (cf.\ (D) and (F)).
}
\label{f:kinetics}
\end{figure}

Fig.~\ref{f:kinetics} (A) and (B) show 
the mean first-passage times (MFPTs) for the binding and unbinding, respectively. 
Note that the BD simulations and FPE calculations 
agree with each other perfectly
for both binding and unbinding, without and with the pocket dry-wet fluctuations, respectively. 
This validates mutually the accuracy of our numerical schemes.
Note also that the binding/unbinding MFPT increases/decreases monotonically as the ligand-pocket
distance increases, due to elongated/shortened ligand travel. 

In Fig.~\ref{f:kinetics} (A), we see that the MFPT for binding is very small if $z<-0.5$ {\AA}.  
This is because the ligand diffusion constant $D_{\rm in}$ inside the pocket is large 
and the PMF is highly attractive; cf.\ Fig.~\ref{f:SolEngy} (B). 
As the initial position $z$ increases from $0$ {\AA} to $5$ {\AA}, 
the difference between the two MFPTs with and without the pocket dry-wet fluctuations
increases from nearly $0$ ps to 
 $100$ ps.
Such an increasing difference results from the existence of the hydration state 2s-wet in 
this range, and the solvation free energy of this state increases as the ligand moves from 
$z = 5 $ {\AA} to $z = 0$ {\AA}; cf.\ Fig.~\ref{f:SolEngy} (A).
The pocket dry-wet fluctuations thus decelerate considerably the ligand-pocket
association.  Such deceleration has been explained by the reduced 
diffusivity of the ligand in the vicinity of pocket entrance 
due to the slow solvent fluctuations~\cite{SMJD_PNAS13}.  
Our predictions of the MFPT for binding, 
with the dry-wet fluctuations included, agree very well 
with the explicit-water MD simulations~\cite{SMJD_PNAS13}, 
improving significantly over those without such fluctuations.  
Note that our model predicts somewhat shorter binding 
times than the MD simulations for $1 < z < 6$~{\AA}. 
In this region, the hydration fluctuations are maximal,
 and this visible but relatively small (when compared to the MFPT from the farthest distance) 
discrepancy reflects some of the approximations of our implicit-solvent theory 
and the model reduction on just a few states.

Fig.~\ref{f:kinetics} (B) 
shows that the timescale for unbinding is significantly larger than that of the binding, by nearly 
three orders of magnitude.  
Without the pocket dry-wet fluctuations, the unbinding MFPT is constant 
for $z < 4$ {\AA} and decreases linearly for $ z > 4$ {\AA}. 
Note that the MFPT for binding in this case also starts to increase significantly at
$z = 4 $ {\AA}; cf.~Fig.~\ref{f:kinetics} (A). 
With the pocket dry-wet fluctuations, the unbinding MFPT is much smaller,  
since the solvation free energy of the 2s-wet state
is higher when the ligand is closer to the pocket (cf.\ Fig.~\ref{f:SolEngy} (A)), 
favoring the ligand unbinding. 
In this case, the MFPT remains constant up to $z = 2 $ {\AA} 
and then decays almost linearly.  This suggests that the wetting transitions 
occur if $z > 2 $ {\AA}.  
Note from Fig.~\ref{f:kinetics} (A) that the 
binding MFPT starts increasing rapidly also around $z = 2$ {\AA}. 

We now study the interesting hydration of the pocket and ligand
individually
during the {\it non-equilibrium} binding/unbinding processes.
For this, we define a pocket hydration parameter to be $ \chi_{\rm p}(z) = 0$ or $1$ 
if the pocket is dry or wet, respectively. Analogously, 
we set for the ligand $ \chi_{\rm l} (z) = 0$ or $1$ 
if the ligand is dry or wet, respectively. 
The values $0$ and $1$ of these ligand-position dependent random variables 
$\chi_{\rm p}(z)$ and $\chi_{l}(z)$ 
are defined by the three hydration 
states 1s-dry, 2s-dry, and 2s-wet (cf.\ Fig.~\ref{f:geometry} (B)--(D)) as follows:  

$\chi_{\rm p} (z) = 0 $ and  $\chi_{l}(z) = 0$  for a 1s-dry VISM surface;   

$\chi_{\rm p} (z) = 0 $ and  $\chi_{l} (z)= 1 $ for a 2s-dry VISM surface;   

$\chi_{\rm p} (z) = 1 $ and  $\chi_{l} (z) = 1 $ for a 2s-wet VISM surface.    

Fig.~\ref{f:kinetics} (C)--(F) show 
the mean values, $\langle \chi_{\rm p}(z) \rangle$ and $\langle \chi_{\rm l}(z) \rangle$, 
and the standard deviations, $\sigma [\chi_{\rm p}(z)] $ and $\sigma [\chi_{ l}(z)], $ 
during the binding and the unbinding processes, respectively.   


When the ligand is far away, there are only two VISM surfaces, 
2s-dry and 2s-wet, cf.\ Fig.~\ref{f:SolEngy} (A).
For such a case, our BD simulations predict the probability 32\% 
of a wet pocket (i.e., $\chi_p=0.32$ for large $z$) in the binding and unbinding processes. 
This is perfectly consistent with the equilibrium probability 
$e^{-G[\Gamma_{\rm{{2s-wet}}}]/k_{\rm B} T} / (e^{-G[\Gamma_{\rm 2s-dry}]/k_{\rm B} T}  
+  e^{-G[\Gamma_{\rm 2s-wet}]/k_{\rm B} T})$ 
predicted by our VISM theory. 
We observe that the pocket hydration peaks at the entrance 
of the pocket in binding, agreeing well with MD simulations~\cite{VISM_LS_PRL09, SMJD_PNAS13},
where it was argued that stronger pocket hydration  is induced by the penetration of the 
ligand solvation shell. When the ligand enters the pocket the latter becomes dry as anticipated.

In comparison, the maximum pocket hydration for unbinding is shifted a bit 
away from the pocket. This kinetic asymmetry or ``translational mismatch'' 
can be explained as well by the asymmetric hydration states of the ligand, 
see Fig.~\ref{f:kinetics} (E), which exits the pocket without a complete solvation shell.
 This behavior is reminiscent of a hysteresis, that is, 
the hydration states during the ligand passage depend on the 
history of the ligand, i.e., where it comes from.

The standard deviations of pocket hydration shown in Fig.~\ref{f:kinetics} (D) depict that 
the dry-wet fluctuations have local maxima close to the pocket entrance 
($z\simeq 3 - 5$~\AA) and behave also significantly different for binding and unbinding.
The corresponding standard deviations of ligand hydration shown in Fig.~\ref{f:kinetics} (F) 
show massively unstable hydration (i.e., large peaks) close to the pocket entrance, 
while inside and far away from the pocket the fluctuations are zero, indicating a very stable 
(de)hydration state. Again the peaks are at
different locations for binding versus unbinding, reflecting the 
hysteresis and memory of dry-wet transitions during ligand passage. 
\section*{Conclusions}
\label{s:Conclusions}

We have developed an implicit-solvent approach, coupling our VISM, 
the string method, and multi-state CTMC BD simulations,  for studying
the kinetics of ligand-receptor binding and unbinding, 
particularly the influence of collective solvent fluctuations on such processes.
Without any explicit descriptions of individual water molecules, 
our predictions of the MFPT for the binding process, which is decelerated by the solvent
fluctuations around the pocket, agree very well with 
the less efficient explicit-water MD simulations. Moreover, we find surprisingly
that the solvent fluctuations accelerate 
the ligand unbinding from the pocket, 
which involves a much larger timescale and is thus more challenging for explicit-water MD simulations 
\cite{ Parrinello_PNAS2015, UbdBerne_PNAS15}.   
Importantly, our implicit-solvent approach indicates that the water effects 
are controlled by a few key physical parameters and mechanisms, such as polymodal 
nano-capillarity based on surface tension of the solute-solvent
interface and the coupling of the random interface forces to the ligand's diffusive motion.

Our approach provides a promising new direction
in efficiently probing the kinetics, and thermodynamics, of the association and 
dissociation of complex ligand-receptor 
systems, which have been studied mostly using enhanced sampling techniques 
\cite{
Shaw_BindingPathway_PNAS2011,
JJBerne_PNAS13,
SMJD_PNAS13,
Parrinello_PNAS2015, 
UbdBerne_PNAS15,
Miao_Biochem2018}. 
Our next step is to extend our approach for more realistic systems
with general reaction coordinates and different techniques for sampling transition paths
\cite{Ciccotti_StringCV_JCP2006,
Stock_CV4ProteinDyn_JCP2018}. 
Our VISM can treat efficiently the electrostatic interactions using the Poisson--Boltzmann 
theory \cite{VISMPB_JCTC2014}.  To account for the flexibility of the ligand and receptor in 
their binding and unbinding, we shall expand our solvation model to include
the solute molecular mechanical interactions  \cite{CXDMCL_JCTC09}.

\section*{Theory and Methods}
\label{s:Theory}


\subsection*{Variational implicit-solvent model (VISM)}
We consider the solvation of solute molecules, with all the solute atomic positions
$\br_1, \dots, \br_N$,  in an aqueous solvent that is treated implicitly as a continuum. 
(For our model ligand-pocket system, the solute atoms include those
of the concave wall and the single atom of the ligand; cf.\ Fig.~\ref{f:geometry}.)
A solute-solvent interface $\Gamma$ is a closed surface that encloses
all the solute atoms but no solvent molecules.  The interior and exterior of $\Gamma$ are 
the solute and solvent regions, denoted  $\Omega_{\rm m}$ 
and $\Omega_{\rm w}$, 
respectively.  We introduce the VISM solvation free-energy functional \cite{DSM06a,DSM06b}: 
\begin{equation}
\label{G}
G[\Gamma] = \Delta P \, \mbox{vol}\, (\Omega_{\rm m}) + \int_\Gamma \gamma \, dS 
+\rho_0  \int_{\Omega_{\rm w}}\!\!\!  U (\br) \, dV + G_{\rm e}[\Gamma].
\end{equation}
Here, $\Delta P$ is the difference of pressures across the interface $\Gamma$, 
$\gamma$ is the solute-solvent interface surface tension, 
$\rho_0 $ is the bulk solvent (i.e., water) density,  
and $U(\br) = \sum_{i=1}^N U_{i} (|\br - \br_i|)$ with each $U_i$ a standard $12$--$6$ LJ potential. 
We take $ \gamma = \gamma_0 (1 -   2 \tau H), $  
where $\gamma_0 $ is the surface tension for a planar interface, 
$\tau$ is the curvature correction coefficient often known as the Tolman length \cite{Tolman49},  
and $H$ is the local mean curvature.  The last term $G_{\rm e}[\Gamma]$ is the electrostatic 
part of the solvation free energy, which we will not include in this study.

Minimizing the functional Eq.~[\ref{G}] among all the solute-solvent interfaces $\Gamma$
determines a stable, equilibrium, solute-solvent interface, called a VISM surface,
and the corresponding solvation free energy.  A VISM surface is termed dry, 
representing a dry hydration state, 
if it loosely wraps up all the solute atoms with enough space for a few solvent molecules, 
or wet, representing a wet hydration state, 
if it tightly wraps up all the solute atoms without extra space for a solvent molecule. 

\subsection*{Implementation by the level-set method} 


Beginning with an initially guessed solute-solvent interface, our level-set method
evolves the interface step by step in the steepest descent direction until a VISM surface is reached.  
Different initial surfaces may lead to different final VISM surfaces.
See Supporting Information (SI) for more details of implementation.


\subsection*{The level-set VISM-string method for minimum energy paths (MEPs)}


Let us fix all the solute atomic positions 
and assume that $\Gamma_0$ and $\Gamma_1$ are two VISM surfaces (e.g., dry and wet surfaces).
We apply the string method~\cite{ERE_JCP07,RE_JCP13} 
to find a MEP that connects $\Gamma_0$ and $\Gamma_1$. 
A string or path here is a family of solute-solvent interfaces $\{ \Gamma_\alpha\}_{\alpha \in [0, 1]}$
that connects the two states  $\Gamma_0$ and $\Gamma_1$.  
Such a string is a MEP, if it is orthogonal to the level surfaces of the VISM free-energy functional. 
To find a MEP connecting $\Gamma_0$ and $\Gamma_1,$ we select some initial images 
(i.e., points of a string), and then update them iteratively to reach a MEP. 
Different initial images may lead to different MEPs. 
Once a MEP is found, we can then find a saddle point on the MEP. 
Alternatively, we can fix one of the VISM surfaces, select some initial images, 
and allow the last image to climb up to reach a saddle point, and then find
the MEP connecting the two VISM surfaces passing the saddle point. 
We refer to SI for more details on our implementation of the method. 

Consider now our ligand-pocket system; cf.\ Fig.~\ref{f:geometry}. 
For any reaction coordinate $z,$  we label all the three 
hydration states 1s-dry, 2s-dry, and 2s-wet (cf.\ Fig.~\ref{f:geometry}) 
as the states $0$, $1$, and $2,$ respectively.  We define for each $i \in \{ 0, 1, 2 \}$ 
the potential
\begin{equation}
\label{Viz}
V_i(z) = G_i(z)+ U_0(z), 
\end{equation}
where $G_i(z)$ is the solvation free energy of the $i$th state at $z$ (cf.\ Fig~\ref{f:SolEngy} (A))
and $U_0(z)$ is the  ligand-pocket vdW interaction potential defined below Eq.~[\ref{V}]. 
We set $V_i(z) = 0$ if the $i$th state does not exist at $z$.

With the energy barriers summarized in Fig.~\ref{f:BarrEngy}, we can calculate for 
each $z$ the rate $R_{ij} = R_{ij}(z)$ of the transition from one state $i$ to another $j.$
If a MEP from $i$ to $j$ passes through another state $k$ (cf.\ Fig.~\ref{f:d6}), 
then we set $R_{ij}(z) = 0.$
If there is only one MEP connecting $i$ and $j$ (see, e.g., $z < 4$ in Fig.~\ref{f:SolEngy}), 
then $R_{ij} = R_0 e^{-B_{ij}(z)/k_{\rm B} T}$ with $B_{ij}(z)$ the 
 energy barrier from $i$ to $j$ and $R_0$ a constant 
 prefactor, describing the intrinsic time scale of water dynamics in the pocket. 
Finally, if there are two MEPs (axisymmetric and axiasymmetric) connecting $i$ and $j$,
we use the same formula but with $B_{ij}$ an effective barrier. 
For instance, consider $i$ and $j$ the states (I) and (IV) in Fig.~\ref{f:d6}, respectively. 
The two transition states are II and III, respectively.  We set 
$B_{ij}(z) = B_{\rm I,IV}(z) = p (G_{\rm II} - G_{\rm I}) + (1 - p) ( G_{\rm III} - G_{\rm I})$, 
where $ p =  e^{-(G_{\rm II}-G_{\rm I}) /k_{\rm B} T}/ 
(e^{-(G_{\rm II}-G_{\rm I}) /k_{\rm B} T} + e^{-(G_{\rm III}-G_{\rm I}) /k_{\rm B} T})$ 
and $G_A$ is the VISM solvation free energy at state $A \in \{ \rm { I, II, III} \}$. 
To determine the prefactor $R_0$, we calculate the equilibrium (i.e., the large $z$ limit) 
energy barriers $B_{\rm dw}$ and $B_{\rm wd}$
in the pocket dry-wet and wet-dry transitions, respectively, 
and equate  $[R_0 (e^{-B_{\rm dw}/k_{\rm B} T} + e^{-B_{\rm wd}/k_{\rm B} T} )]^{-1}$
with the time scale for the relaxation of water fluctuation  of $10$ ps 
as predicted by explicit-water MD simulations \cite{SMJD_PNAS13}. 
See SI for discussions on the sensitivity of the results on $R_0.$ 

\subsection*{Continuous-time Markov chain (CTMC) Brownian dynamics (BD) simulations 
and the mean first-passage time (MFPT)}

To include explicitly the dry-wet fluctuations, we introduce a position-dependent, multi-state,
random variable $\eta = \eta(z)$:  
$\eta(z) = i$ $(i\in \{ 0, 1, 2\})$ if the system is in the $i$th hydration state
when the ligand is located at $z$, with the transition rates $R_{ij} (z)$ given above. 
We define the potential $V_{\rm fluc}(\eta, z) = V_{i}(z) $ (cf.\ Eq.~[\ref{Viz}]) 
if $\eta(z) = i.$ \cite{Doering_PRL92}.  
The random position $z = z(t)  = z_t$ of the ligand is now determined 
by our CTMC BD simulations in which we solve the stochastic differential equation 
\[
\begin{aligned}
&dz_t = \left[ - \frac{D(z_t)}{k_{\rm B} T} \frac{ \partial V_{\rm fluc} 
(\eta(z_t), z_t)}{\partial z} +  D'(z_t) \right]  dt + \sqrt{2 D(z_t)} \, d \xi_t.  
\end{aligned}
\]
Here, the partial derivative of $V_{\rm fluc}$ is with respect to its second variable, 
$D(z)$ is an effective diffusion coefficient that smoothly interpolates
the diffusion coefficients $D_{\rm in}$ and $D_{\rm out}$ inside and outside the pocket, 
respectively, and 
$\xi_t $ is the standard Brownian motion.  Solutions to this equation are constrained by
$z_t \in [z_{\rm L}, z_{\rm R}] $ for some $z_{\rm L}$ and $z_{\rm R}.$
For the simulation of a binding process, we reset the value of $z_t $ to be $2z_{\rm R} - z_t $ 
if $z_t \ge  z_{\rm R}$, and we stop the simulation if $z_t  \le z_{\rm L}$. 
For the simulation of an unbinding process, we reset the value of $z(t)$ to be $z_{\rm L}$ if
$z_t \le z_{\rm L}$, and we stop the simulation if $z_t \ge z_{\rm R}.$   
The distribution of $\eta(z_0)$ for an initial ligand position $z_0$ 
is set based on the equilibrium probabilities 
${e^{-G_i/k_{\rm B}T }}/{\sum_{j=0}^2 e^{-G_j/k_{\rm B}T } } $ $(i=0, 1, 2),$ 
where $G_i$ is the solvation free energy of the $i$th hydration state at $z_0.$ 

We run our CTMC BD simulation for the ligand starting at a position $z_0 = z_{\rm init}$, 
and record the time at which the ligand reaches $z_{\rm L}$ (or $z_{\rm R}$) for the first time
for a binding (or unbinding) simulation. 
We run simulations for $3,000$ times and average these times 
to obtain the corresponding MFPTs.

\subsection*{Fokker--Planck equations (FPE) and the MFPT}

The probability densities $P_i = P_i (z,t)$ for the ligand at location $z$ at time $t$ with 
the system in the $i$th hydration state are determined by the generalized FPEs 
\cite{Doering_PRL92, JJBerne_PNAS13}: 
\[
\begin{aligned}
\frac{\partial P_{i}}{\partial t} &= 
\frac{\partial}{\partial z}\left\{ D(z) \left[ \frac{\partial P_{i}}{\partial z}
+ \frac{1}{k_{\rm B} T }   V'_{i} (z) P_i \right] \right\} \nonumber \\
&\quad +\sum_{0\le j \le 2, j\not =i} R_{ji}(z)  P_j 
-\biggl(\, \sum_{0\le j \le 2, j\not =i} R_{ij}(z) \biggr) P_i 
\end{aligned}
\]
for $i =0, 1, 2,$ where $V_i$ is defined in Eq.~[\ref{Viz}].
These equations are solved for $z_{\rm L} < z < z_{\rm R},$ with the boundary conditions
$P_{i}(z_{\rm L}, t) =0$ and $ \partial_z P_{i}(z_{\rm R},t) =0$ for binding,
and $\partial_z P_{i} (z_{\rm L},t)   + ({1}/{k_{\rm B}T}) 
V'_{i}(z_{L}) P_{i} (z_{\rm L}, t)  =0$ and  $P_{i} (z_{\rm R}, t) =0$ 
for unbinding, respectively.
The initial conditions are
$P_i (z,0) = \delta (z - z_{\rm init})$ if the ligand is initially at $z_{\rm init}.$
We obtain the MFPT as the double integral of $\sum_{i=0}^2 P_i(z,t)$
over $(z, t) \in [z_{\rm L}, z_{\rm R}]\times [0, \infty).$


\subsection*{Parameters} 

We set the temperature $T = 298$ K, bulk water density $\rho_0 = 0.033 \ \mbox{\AA}^{-3}$,  
the solute-water surface tension constant $\gamma_0 = 0.143 \ k_{\rm B} T/\mbox{{\AA}}^2$
($k_{\rm B}$ is the Boltzmann constant),  and the Tolman length $\tau = 0.8$~{\AA}.
We set $\Delta P \mbox{vol}\,(\Omega_{\rm m}) = 0$ as it is relatively very small. 
The LJ parameters for the wall particles, ligand, and water are 
$\varepsilon_{\rm wall} = 0.000967~k_{\rm B} T$ and $\sigma_{\rm wall} = 4.152$~\AA,  
$\varepsilon_{\rm ligand} = 0.5 \ k_{\rm B} T$ and $\sigma_{\rm ligand} =3.73$~\AA, 
and $\ve_{\rm water} = 0.26 \ k_{\rm B} T$ and $\sigma_{\rm water} = 3.154$~{\AA}, respectively.  
The interaction LJ parameters are determined by  the Lorentz--Berthelot mixing rules. 
The prefactor $R_0 = 0.13$ ps$^{-1}$.
The diffusion constants are  $D_{\rm out} = 0.26 \ \mbox{\AA}^2$/ps 
\cite{SMJD_PNAS13}, 
and $D_{\rm in} = 1 \ \mbox{\AA}^2$/ps. 
The cut-off position distinguishing the inside and outside of the pocket is $z_{\rm c} = -0.5$~{\AA}.
BD simulations and FPE calculations are done for $z_{\rm L} \le z \le z_{\rm R}$ 
with $z_{\rm L} = - 4$~{\AA} 
and $z_{\rm R} = 15.5$~{\AA}.




\acknow{
SZ was supported in part by NSF of Jiangsu Province, China, through grant BK20160302, NSFC through grant NSFC 21773165 and NSFC 11601361, and Soochow University through a start-up grant Q410700415.  RGW and JD thank the DFG for financial support.  JD also acknowledges funding from the ERC within the Consolidator Grant with Project No.~646659--NANOREACTOR.  Work in the McCammon group is supported in part by NIH, NBCR, and SDSC.  LTC and BL were supported in part by the NSF 
through the grant DMS-1620487. SZ thanks Dr.\ Yanan Zhang for helpful discussions on the string method.
}

\showacknow{} 

\bibliography{hydrophob}
\end{document}